# Synthesis of A Novel Super Absorbent Based on Wheat Straw

Anupam Saha[1,*], Md. Shahjahan[2]

[1] Department of Materials Science and Engineering, Khulna University of Engineering & Technology, Khulna-9203, BANGLADESH

[2] Department of Electrical and Electronic Engineering, Khulna University of Engineering & Technology, Khulna-9203, BANGLADESH

E-mail: anupam.kuet.mse@gmail.com

## Abstract

*The purpose of this research is to generate wheat straw based super absorbent which is eco-friendly. The chemical approach of graft-copolymerization is used to create this novel super absorbent polymer. Polyacrylic acid is the building block polymer that is utilized in the production of highly absorbent polymer. Furthermore, water absorption capabilities of these environmentally friendly absorbents are investigated in a variety of conditions and then compared to one another. The polymer is then tested for absorbency. It is discovered that several optimized factors, such as the time span of the immersed sample, the effect of different solutions, and the temperature, has an impact on the absorbencies of the super absorbent. Optical microscope pictures provided proof of the morphological characteristics. The excellent product is achieved with absorbencies of 251 percent in distilled water and 528.34 percent in a 4-weight percent sodium hydroxide solution, respectively. It is also discovered that the absorbency increased in accordance with the increase in temperature. At 80 degrees Celsius, the maximum absorbency is observed in water, at 548.6 percent.*

## 1. Introduction

Superabsorbent polymers (SAPs), also known as hydrogels or hydrocolloids, can absorb and hold many times their own weight in fluids like water under moderate pressure. A change in the size of polymer gels as a result of exposure to particular environmental conditions alters their core structure as well as the optical and mechanical characteristics of the polymer network. It is because of this ability to withstand and absorb fluids up to a hundred times their own mass that they are referred to as "superabsorbent materials. "Superabsorbent polymers are now often employed in the production of sanitary items that absorb and retain bodily fluids [1]. The water absorption capacity (WAC) of superabsorbent polymers is the most essential feature of these materials. Its water absorption coefficient (WAC) is determined by the composition and structure formed by the production technique, as well as by the presence of electrolytes in the water source. The first introducing of SAP is when acrylic acid (AA) and divinylbenzene undergoes thermal polymerization in an aqueous medium in late 1983s. Then, SAP was developed into new generation of SAP which hydrogels in era 1950s. In year 1970s, the first commercial. SP was introduced by undergoes alkaline hydrolysis process of starch-graft-polyacrylonitrile (SPAN). Then, after 8 years of SPAN was introduced by Zohuriaan and Kabiri in 2008, Japan developed more SAP materials to be used in their feminine industry [2]. Various functions of SAP were demonstrated by Zohuriaan and Kabiri include soil additions to enhance physical qualities of soil, macro-porous medium, and retaining materials. SAP may also be employed as a controlled release system by absorption of certain nutrient components from the soil [3]. Because of the evolution of superabsorbent polymer composites to Nano-scale particles, its application as a high-performance engineering material has increased. Chitosan is another significant natural polymer that has been changed chemically, for example, by reacting with vinyl monomers and APS to form the superabsorbent polymer carboxymethylchitosan (CMCTS-g-PAA). Thus, this SAPC demonstrated that the rate of water absorption by polymer composites was high, as Yu and Hui-min discovered. This demonstrates that by adding filler to the SAPC system, a high level of water absorbency is achieved for this research [4].In Bangladesh, Wheat and Rice are the main crops produced. The straws of these crops are considered as waste materials. Millions of wheat straws are wasted here yearly. So, production of superabsorbent based on these straws can be beneficial in both ways for Bangladesh. One, wheat straws wastages can be minimized. Two, Biodegradable superabsorbent can be useful to the applications of absorbent polymer. Moreover, this superabsorbent can used in many applications in the context of Bangladesh.

Wheat straw can be modified through chemical method and used as skeletal material, on which some monomers such as AA and AM can graft so as to form superabsorbent composite [5]. In the last several years, significant progress has been achieved in this subject. Using the copolymerization of wheat straw and acrylic acid, researchers were able to create a single ion SAR [6]. To develop the super-absorbent composite, wheat straw was chemically treated to form cellulose derivatives such as carboxymethyl cellulose, which was then grafted with certain monomers to make the super-absorbent composite [7]. But there are few reports on the preparation of an amphoteric super-absorbent polymer, which have better water absorbency than the single-ion SAR by using wheat straw directly [8]. But no researches have showed the use of this biodegradable absorbent in real life practice and how it will compete with other available super absorbent in terms of properties. So, in this work, A novel super-absorbent will be synthesized based on wheat straw nanoparticles which is an organic compound and this will be biodegradable. And absorbency of it in different solutions will be checked.

## 2 Methodology and Materials

### Methodology
Total methodology of this experiment can be divided into these parts:

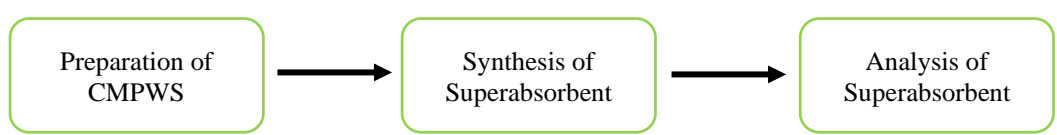

### Material Selection
Acrylic Acid ($C_3H_4O_2$) of reagent grade purity 98.5%, Ammonium per sulfate (APS), N-N.-Methylene-Bis-Acrylamide grade purity 97% were purchased from Khulna scientific, Khulna, manufactured, by Avon Chemical, United Kingdom. Sodium hydroxide (NaOH) solution and Chloroacetic acid were used for pulverization. Brine water solution, NaOH solution and Saline solution were made in the laboratory to do the absorbency test of the superabsorbent. Organic wheat straw was bought from Panchagarh district. It is an attractive class of organic compound used for synthesis of water absorbent.

### Process of making CMPWS
The wheat straws are chopped and dried in an $80^0$ C drying oven for 8 h. The dried wheat straws are then grinded and the fine wheat straw powder is weighed 40g and dipped in a 500 ml beaker with 4% sodium hydroxide solution. The aqueous suspension is heated and stirred at $90^0$ C for 3 h, and then the aqueous suspension filtered and washed with 95% ethanol solution. The filter residue is transferred into a beaker with 2.5% sodium hydroxide solution in water bath, and then interfused by 5g chloroacetic acid, and being stirred and heated at $70^0$ C for 1 h. After that, the mixed solution is filtered with 75% ethanol solution, and the chemically modified pulverized wheat straw (CMPWS) is dried at $150^0$C.

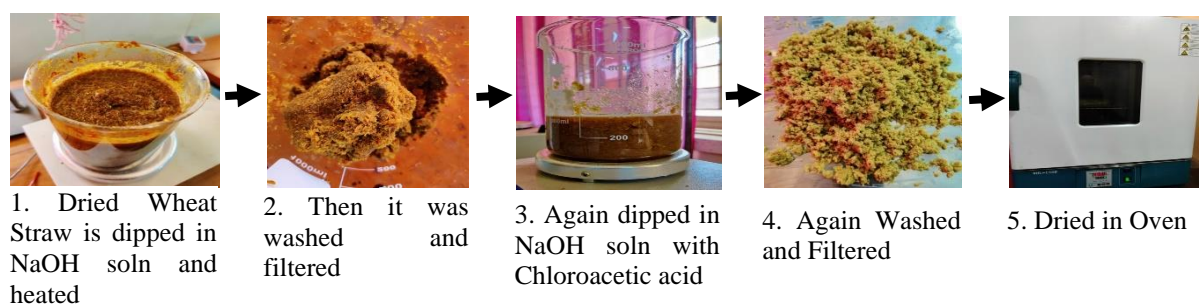

1. Dried Wheat Straw is dipped in NaOH soln and heated
2. Then it was washed and filtered
3. Again dipped in NaOH soln with Chloroacetic acid
4. Again Washed and Filtered
5. Dried in Oven

**Fig. 2.1.** Process of making Pulverized Wheat Straw

### Preparation of Super-Absorbent Polymer based on CMPWS
Dried Chemically modified pulverized wheat straw was grinded and gone through sieve to make Nano particles powder. An amount of chemically modified pulverized wheat straw powder (CMPWS), N, N'-methylenebisacrylamide (MBA), ammonium per sulfate (APS) and acrylic acid (AA) with different degrees of neutralization were prepared by the following procedure. A weight quantity of dried CMPWS powder and 20g Acrylic acid were added into a flask and stirred. A solution of 5g ammonium per sulfate (APS) after neutralization, certain amount (1g) of MBA and 10ml de-ionized water were added. The water bath was kept at high temperature for 1.5h to complete polymerization. The resulting product was then dried at constant 70$^0$C for 1h to get the sol gel product.

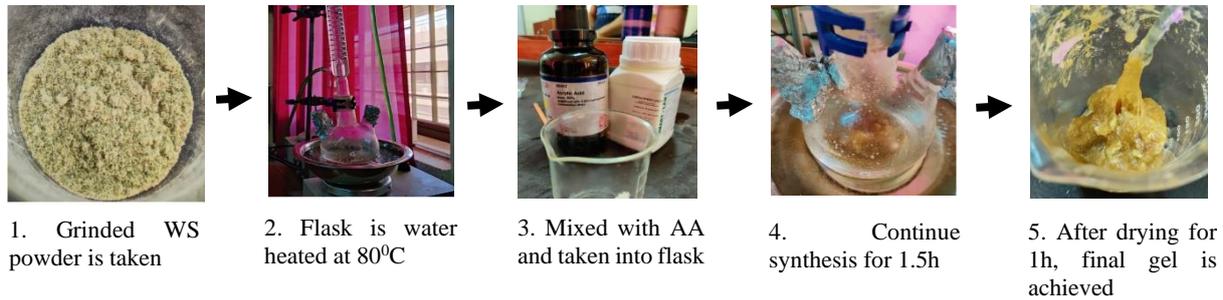

1. Grinded WS powder is taken
2. Flask is water heated at 80$^0$C
3. Mixed with AA and taken into flask
4. Continue synthesis for 1.5h
5. After drying for 1h, final gel is achieved

**Fig. 2.2.** Process of making Super-absorbent solgel

**Measurement of water absorbance**
A specific quantity of the super-absorbent was immersed in distilled water, brine water, NaOH aqueous solution and saline solution to swell equilibrium, respectively. Swollen samples were then separated from unabsorbed water by filtered over a sieve screen. The water absorbency ($Q_{H2O}$) of super-absorbent composite was determined by weighing the swollen samples and calculated using the following equation:

$$Q_{H2O} = \frac{m2-m1}{m1} * 100\% \quad (1)$$

m1 and m2 are the weights of the dry sample and the water-swelling sample (g), respectively. $Q_{H2O}$ is expressed as grams of water per gram of sample (g/g). Then it is mixed with distilled water and then allowed to hydrate at different temperatures to determine the absorption rate at different temperatures. And finally, these results are analyzed.

## 3 Results and Discussion

**Influence of Different Solutions on Absorbency of Superabsorbent based on Wheat Straw**

Superabsorbent polymers were immersed in brine water, NaOH solution and saline solutions in order to test their ability to absorb water. Water absorption was studied on SAPC samples. Although the weight proportion of cross-linker and initiator remained consistent, the solution in each sample was different. Water absorption was measured by examining samples of superabsorbent polymer in various solutions for varied amounts of time.
Water absorption by the specimen containing 54% Wheat straw was measured at 147 percent after it was submerged in brine water for one hour. However, specimens submerged in brine water showed an increase in water absorption as the immersion period increased. After two hours in brine water, the sample had the maximum water absorbency of 194 percent. Increase of absorbency continued until 4 hours. Then it reached saturation point. And the maximum absorbency was 215 in Brine water.

The water absorbency of a specimen that had been soaked in water for one hour was measured to be in the region of 205 %. However, it was shown that increasing the duration spent submerged in water resulted in an increase in the water absorbency. After 2 hour the absorbency was 251%. Water absorbency in water reached saturation point after Six hours and the maximum absorbency was 503%.

When immersed in 4% NaOH solution, polymer absorbed more water than other solutions. Water absorbency by the superabsorbent immersed for one hour in NaOH solution was measured in the range of 477%. However, it was also found that as the time of immersion was increased, increase in water absorbency for specimens immersed in NaOH solution was witnessed. The water absorbency of 528.34% was measured in the specimen having 54% wheat straw immersed in NaOH solution for Two hours. Increasing of absorbency continued till 10-12 hours for NaOH solution. And maximum absorbency was 1236%.

The water absorbency of the superabsorbent after being submerged in Saline solution for one hour was measured to be in the region of 357 %. However, it was shown that increasing the length of immersion for specimens immersed in Saline solution resulted in an increase in the water absorbency of the specimens examined. SAP soaked in a saline solution for two hours had with a reading of 405.7 %. It reached saturation point after 10 hours. Maximum absorbency in saline solution was 1040%.

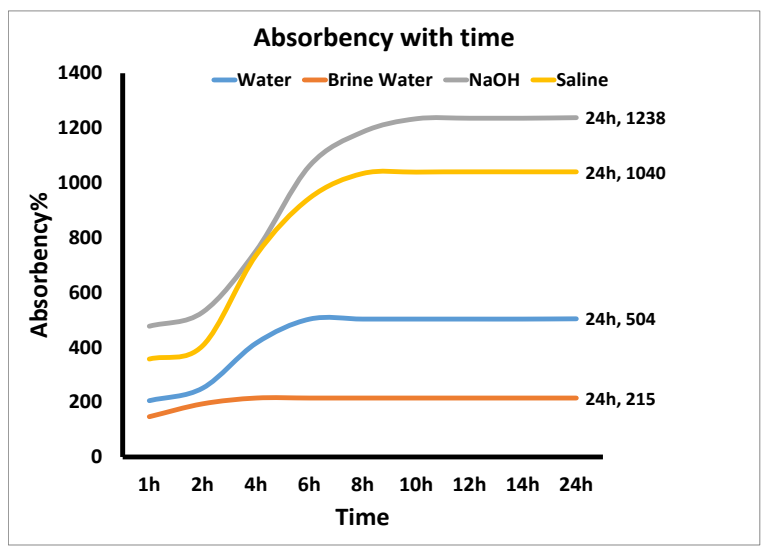

**Fig. 3.1.** Absorbency with increasing time span for different solutions at room temperature

Fig. 3.1 shows the graphical representation of absorbency in water, brine water, NaOH and Saline solutions. It is seen that for brine solution it is the least and for NaOH solution it is the most.

For brine solution, the interactions between wheat straw and acrylic acid decreased gradually with increasing time consumption, resulting in decreased creation of physical and chemical cross-links in the matrix of polymer, and the elasticity of the polymer chains increases with more crosslinks in the polymer composite.

The ability of the superabsorbent polymer composite to absorb NaOH solution is greater than water, saline water or brine solution. Carbon dioxide capture by chemical absorption has been one of the most popular approaches. However, low solubility of $CO_2$ in water and high energy demand for absorbent regeneration hinder its feasibility in water. When NaOH was added with water these drawbacks were hindered and it consumed more water.

**Effect of Temperature on Absorbency**

To measure the effect of temperature in absorbency, samples were dipped in water solution at $28.3^0C$, $40^0C$, $60^0C$ and $80^0C$ respectively. The water absorbency of the specimen in water at $28.3^0C$ was observed by keeping 1 hour. It was seen that with time water absorbency was increasing. At $28.3^0C$ the water absorbency was in the range of 251%. Water absorbency seemed to increase with increasing temperature. Specimen was kept in water at $40^0C$ for 1 hour. The water absorbency at $40^0C$ was in the range of 368.2%. So, it was seen temperature increasing effecting the water absorbency. In the case of $61.3^0C$, Water absorbency appeared to rise more rapidly as temperature increased. For one hour, the specimen was maintained in water at $61.3^0C$. At $61.3^0C$, the water absorbency was in the region of 408.6 %. Temperature increases resulted in an increase in absorbency. The absorbent polymer in aqueous solution exhibited the greatest absorbency when heated to $80^0$ Celsius. For one hour, the specimen was submerged in water at 80 degrees Celsius. At 80 degrees Celsius, the water absorbency was in the region of 548.6 %. As a result, the influence of temperature was thoroughly investigated, and it was discovered that the absorbency rose as temperature increased.

Fig 3.2 shows that absorbency of the superabsorbent specimen increases with increasing temperature. At $28.3^0C$ it was 251, at $40^0C$ it was 368.2, at $61.3^0C$ it was 408.6 and at $80^0C$ it was 548.6.

Increased molecule movement (vibration and rotation) occurs at higher temperatures; hence, more energy is required at higher temperatures, resulting in a rise in absorbance. The interactions between wheat straw and acrylic acid decreased gradually with increasing temperature, resulting in decreased cross-linking of the polymer matrix in a physical and chemical manner, which enhances the elasticity of the polymer chains. That's why absorbency increases with increasing temperature.

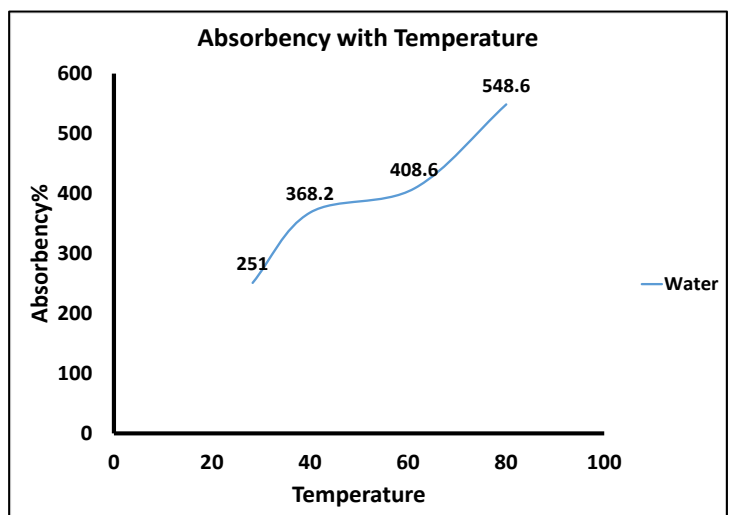

**Fig. 3.2**. Absorbency with increasing temperature in Water

**Microstructure of Wheat Straw based Superabsorbent**

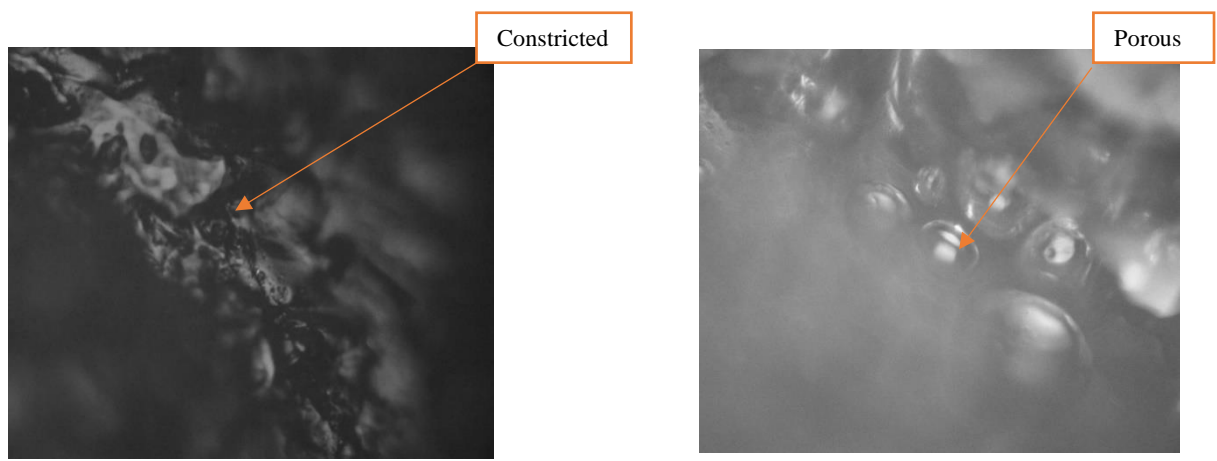

1) Constricted Surface at 28.3$^0$C       2) Porous Surface at 28.3$^0$C

**Fig. 3.3.** Microstructure of surface of Wheat Straw based Superabsorbent measured by optical microscope.

Fig 3.3 shows microstructures taken with an optical microscope of the pulverized wheat straw samples that have been treated with acrylic acid. At the lower temperature of 28.3$^0$C (Left, Fig. 3.3), a tighter surface and less porous structure can be detected, but a larger network and more porous structure increase the surface area of the copolymer hydrogel at the higher temperature of 80$^0$C (Right, Fig. 3.3). While immersed in water, the gel mass efficiently diffused the aqueous medium through the micro porosity spaces, resulting in a significant increase in the absorption rate of the solution. Therefore, the temperature had a significant impact on the surface morphology of the super-absorbent material. Interestingly, this discovery was in excellent accord with our water absorption measurements.

## 4 Conclusion
From the results, the following conclusions can be deducted:

By altering the solutions, the water absorbency of the superabsorbent polymer may be enhanced. The absorbency of brine water was lower, but the absorbency of NaOH solution was higher. As a result, it can be stated that the NaOH solution containing 54 weight percent of Wheat straw in the polymer matrix provides the highest water absorbency. The amount of water that can be absorbed depends on the amount of time that has passed. It rises in proportion to time, such that the best absorbency may be reached at a longer time interval. Increasing the temperature causes the water absorbency of the polymer to rise, and because of the reduced proportion of cross-linker in the polymer, a porous structure is created, resulting in an increase in the porosity of the super absorbent polymer composite. For lower temperature surface of the superabsorbent was tighter and less porosity was seen. When temperature increases, the surface became porous and more absorbency was seen.